\documentclass[american,10pt, journal, twocolumn]{paper}
\usepackage[T1]{fontenc}
\usepackage[utf8]{inputenc}
\usepackage{varwidth}
\usepackage{amsmath}
\usepackage{amssymb}
\usepackage{graphicx}

\makeatletter

\providecommand{\tabularnewline}{\\}
\newenvironment{cellvarwidth}[1][t]
    {\begin{varwidth}[#1]{\linewidth}}
    {\@finalstrut\@arstrutbox\end{varwidth}}

\let\paperclassexample\example

\let\example\relax
\AtBeginDocument{%
  \@ifundefined{example}{\let\example\paperclassexample}{}
}

\usepackage{balance}
\usepackage{spconf}
\usepackage[pdftex,pdftitle={OLAC: An Overlapped Lossless Audio Codec in the Time-Domain with MDCT Compatibility},pdfauthor={Jean-Marc Valin}]{hyperref}
\ninept

\makeatother

\usepackage{babel}
\begin{document}
\title{OLAC: An Overlapped Lossless Audio Codec in the Time-Domain\\ with MDCT Compatibility}
\name{Jean-Marc Valin}
\address{Google LLC}
\maketitle
\begin{abstract}
Lossless audio coding is a highly mature field of research, with limited
potential for significant improvements in pure compression performance.
However, emerging real-time wireless applications increasingly require
dynamic transitions between lossy and lossless coding to adapt to
fluctuating network capacities. Existing standalone lossless codecs
cannot achieve this seamless switching without discontinuity. In this
paper, we propose a lossless codec based on time-domain aliasing cancellation
(TDAC) that can match the overlap in the CELT mode of the Opus codec.
This allows the transition between lossy and lossless coding to be
achieved without discontinuity or the transmission of redundant information.
We show that the proposed codec still achieves state-of-the-art lossless
compression without being penalized by its use of TDAC. 
\end{abstract}

\section{Introduction}

There are currently many available standalone lossless audio coding
formats. These include FLAC~\cite{van2024rfc,flac}, ALAC~\cite{alac},
MPEG-4 ALS~\cite{liebchen2004mpeg}, IEEE~1857.2~\cite{huang2014lossless},
and Monkey's Audio~\cite{monkeysaudio}. The majority of lossless
audio codecs operate in the time domain, using some form of linear
prediction~\cite{makhoul1975linear}. One of the few exceptions is
MPEG's Scalable to Lossless (MPEG-4 SLS)~\cite{yu2004mpeg} scheme,
which is based on a perfectly reversible integer MDCT (IntMDCT)~\cite{geiger2001audio}.
While the IntMDCT makes it possible to produce an embedded codec using
an AAC core, it also introduces some costs, both in terms of code/computational
complexity and performance~\cite{huang2014lossless}. 

In emerging audio applications, seamless switching between lossy transform
coding and lossless coding is highly desirable. For example, wireless
audio protocols attempt lossless transmission over a fixed-bandwidth
channel, but require an instantaneous, glitch-free fallback to lossy
coding when wireless interference reduces channel capacity. Similarly,
when performing lossy coding at high bitrates ($>200$ kb/s per channel),
opportunistic lossless coding can seamlessly save bits when the audio
complexity is low enough to allow perfect reconstruction at a lower
bitrate than the lossy option.

Unfortunately, switching between lossy and lossless with existing
time-domain audio codecs without introducing discontinuities involves
coding redundant information both to cover the MDCT overlap and to
avoid the abrupt change in quantization noise. Switching would be
achievable with the IntMDCT by using the same window as the lossy
codec, but this sacrifices the simplicity of time-domain lossless
coding. 

In this paper, we propose an overlapped lossless audio codec (OLAC)
that integrates seamlessly with the CELT~\cite{valin2013high} transform
coding mode of the Opus~\cite{rfc6716} codec. Our primary contribution
is a novel scheme that achieves MDCT-compatible time-domain aliasing
cancellation (TDAC) entirely within the time domain, using only linear
prediction (Section~\ref{sec:OLAC-Overview}). By deliberately omitting
the frequency-domain transform step required by integer MDCT schemes,
our approach maintains the simplicity and low computational overhead
of traditional time-domain lossless coding while enabling perfect,
discontinuity-free switching to and from lossy transform modes (Section~\ref{sec:Integration-and-Switching}).
We further show in Section~\ref{sec:Evaluation} that OLAC achieves
state-of-the-art lossless compression without being penalized by its
use of TDAC, even for short frame sizes.

\section{OLAC Overview}

\label{sec:OLAC-Overview}

The core architecture of OLAC is designed to perfectly mirror the
overlap-add structure of the CELT lossy codec while operating entirely
in the time domain. At a high level, the OLAC encoder processes audio
through three sequential stages: 1) a reversible TDAC windowing step
applied only to the overlap regions to ensure MDCT compatibility,
2) forward-adaptive linear prediction to decorrelate the time-domain
signal, and 3) entropy coding of the residual using Golomb-Rice codes.
To match CELT's signal conditioning, a perfectly reversible pre-emphasis
filter is also applied. 

\subsection{Time-Domain Aliasing Cancellation (TDAC)}

\begin{figure}
\centering{}\includegraphics[width=1\columnwidth]{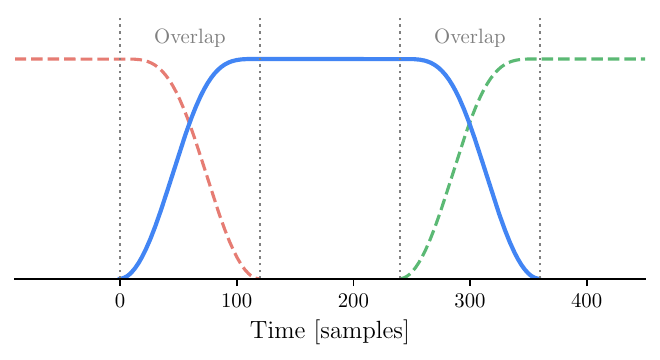}\caption{Windowing used in CELT and OLAC for the case of a 240-sample (5-ms)
frame size. The overlap is always 120~samples (2.5~ms), regardless
of the frame size.}\label{fig:Windowing-CELT-OLAC}
\end{figure}

\begin{figure}[t]
\centering{}\includegraphics[width=1\columnwidth]{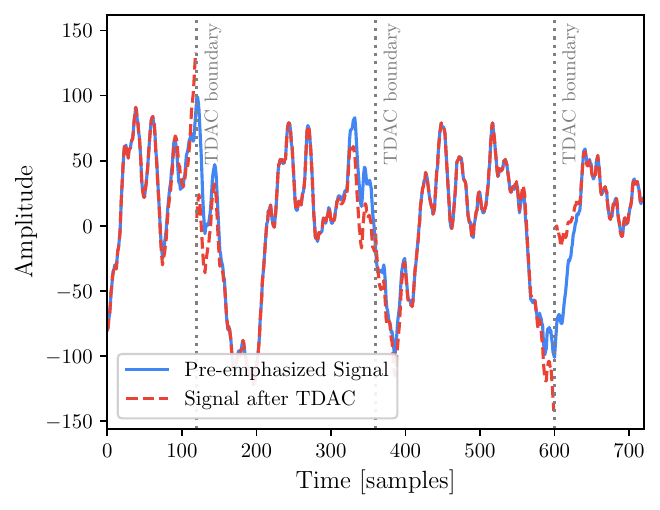}\caption{Illustration of the effect of TDAC windowing on the pre-emphasized
signal. To make the effect more obvious, we use a shorter frame size
of 240~samples (5~ms). We can observe that the signal becomes discontinuous
at the frame boundaries, but over a given frame, the signal is still
smooth.}\label{fig:Illustrating-TDAC}
\end{figure}

The modified discrete cosine transform (MDCT)~\cite{malvar1990lapped}
is based on the principle of time-domain aliasing cancellation~\cite{princen1986analysis}.
The MDCT is typically implemented in two steps. First, a TDAC step
is performed using a complementary window. Second, a DCT-IV converts
the signal to the frequency domain. In~\cite{geiger2001audio}, both
steps are implemented using a perfectly reversible integer lifting
transform~\cite{daubechies1998factoring}. In this work, we skip
the second step and perform only the windowing step, applying the
Vorbis~\cite{vorbis_spec} window used in CELT: 
\begin{equation}
w\left(n\right)=\sin\left[\frac{\pi}{2}\sin^{2}\left(\frac{\left(n+1/2\right)\pi}{2L}\right)\right]\,,\label{eq:vorbis_window}
\end{equation}
where $L$ is the length of the overlap. Since CELT uses a flat-top
window (Fig.~\ref{fig:Windowing-CELT-OLAC}), TDAC is only applied
to the overlap region. For that reason, OLAC look-ahead is equal to
the overlap duration (2.5~ms) and is the same as for CELT.

Let $r(n)$ denote the overlap region of length $L$. For each pair
of points on either side of the mid-point of the overlap, the overlapped
TDAC signal can be expressed as:
\begin{equation}
\left[\begin{array}{c}
s\left(n\right)\\
-s\left(L-n-1\right)
\end{array}\right]=W\left[\begin{array}{c}
r\left(L-n-1\right)\\
r\left(n\right)
\end{array}\right]\,,\label{eq:TDAC_def}
\end{equation}
where
\begin{equation}
W=\left[\begin{array}{cc}
w\left(n\right) & w\left(L-n-1\right)\\
-w\left(L-n-1\right) & w\left(n\right)
\end{array}\right]\,.\label{eq:TDAC_W}
\end{equation}
Since perfect reconstruction requires $w^{2}\left(n\right)+w^{2}\left(L-n-1\right)=1$,
$W$ is a rotation matrix. To implement the TDAC rotation using the
standard lifting factorization in Eq.~(7.2) of~\cite{daubechies1998factoring},
we define the rotation angle $\alpha$ such that $\cos\alpha=w\left(n\right)$
and $\sin\alpha=-w\left(L-n-1\right)$. Substituting these into the
standard lifting steps results in 
\begin{equation}
W=\left[\begin{array}{cc}
1 & p\\
0 & 1
\end{array}\right]\left[\begin{array}{cc}
1 & 0\\
q & 1
\end{array}\right]\left[\begin{array}{cc}
1 & p\\
0 & 1
\end{array}\right]\,,\label{eq:lifting_factorization}
\end{equation}
where $p=\left(1-w\left(n\right)\right)/w\left(L-n-1\right)$ and
$q=-w\left(L-n-1\right)$. 

Using integer rounding, we can express the reversible TDAC as
\begin{align}
a & \leftarrow r\left(L-n-1\right)+\left\lfloor pr\left(n\right)\right\rceil \\
b & \leftarrow r\left(n\right)+\left\lfloor qa\right\rceil \label{eq:lifting}\\
s\left(n\right) & \leftarrow a+\left\lfloor pb\right\rceil \\
s\left(L-n-1\right) & \leftarrow-b\,,
\end{align}
where $\left\lfloor \cdot\right\rceil $ denotes rounding to the nearest
integer. Fig.~\ref{fig:Illustrating-TDAC} illustrates how the signal
is modified by the TDAC process.

\subsection{Prediction}

OLAC uses forward prediction, estimating the optimal prediction filter
using Burg spectral estimation~\cite{burg1967} with order $P_{\mathrm{max}}=63$.
Since very high-order filters are rarely beneficial, we truncate the
filter to order $P$ by minimizing the total rate
\begin{equation}
P=\underset{p}{\mathrm{argmin}}\left(pR-\frac{L_{\mathrm{frame}}}{2}\log_{2}G_{p}\right)\,,\label{eq:Burg-truncation}
\end{equation}
where $G_{p}$ is the prediction gain when truncating at order $p$,
$L_{\mathrm{frame}}$ is the frame size, and $R=3.5$ bits/coefficient
is the approximate rate for coding each coefficient. The optimal order
$P$ varies from 0 to $P_{\mathrm{max}}$, but typically averages
around $P=30$.

The process of applying TDAC results in a discontinuity between the
end of a frame and the beginning of the next frame. This prevents
the optional cross-frame prediction feature available in some other
lossless codecs. For this reason, along with the relatively small
frame size we use (20~ms or less), it is important to minimize the
unpredicted samples at the beginning. Whereas FLAC directly codes
the first $D$ samples when using a filter of order $D$, we adopt
the progressive-order prediction proposed in~\cite{moriya2004extended}.
By representing the prediction filter in the reflection coefficient
(PARCOR) domain, the conversion to direct-form coefficients produces
filters for all orders up to $D$. As a result, only the first residual
sample of the frame is completely unpredicted. 

To make implementation easier on low-power devices, we avoid the use
of 64-bit accumulators in the prediction. This is achieved by adaptively
scaling down the filter coefficients and signal based on their maximum
magnitude to 13~bits and 12~bits, respectively. In the case of the
prediction coefficients, the encoder and decoder each compute the
shift independently based on the decoded coefficients, so no shift
information needs to be coded. Conversely, the signal shift is computed
by the encoder and transmitted in the bitstream. The truncated signal
least significant bits (LSBs) are then coded directly with no prediction
or entropy coding. As a result, the encoder and decoder predictors
can operate using 16-bit multiply-accumulate (MAC) operations. Given
the maximum order of $P_{\mathrm{max}}=63$, the resulting accumulation
is guaranteed to fit a 32-bit integer. That is the case even for 24-bit
audio. 

\subsubsection{Coefficient Coding}

It is common to code reflection coefficients in a warped domain, such
as log area ratio (LAR) or arcsin~\cite{liebchen2004mpeg}. For lossy
speech coding, such quantization results in a more accurate spectral
representation. However, in the context of lossless audio coding,
the criterion that ultimately matters is how the mean squared error
of the direct-form coefficients (which is proportional to the prediction
error caused by quantizing the filter) relates to the prediction error
of the unquantized filter. For that reason, we adjust the quantization
resolution based on the gain of the prediction filter. Also, not all
reflection coefficients are equally important. Ideally, the resolution
of each coefficient should correspond to the prediction gain of a
filter that excludes all previously coded reflection coefficients. 

Each coefficient is quantized as $\hat{k}_{n}=\left\lfloor k_{n}Q_{n}\right\rceil /Q_{n}$.
Based on the prediction gain of the entire filter $G_{P}$, we select
the number of quantization steps to use for $k_{0}$ as $Q_{0}=K_{q}\sqrt{G_{P}}$,
where $K_{q}=9$ is a tunable encoder-only parameter. The value of
$Q_{0}$ is rounded up to the next power of two and transmitted in
the bitstream.

As we decode coefficients, we remove the contribution of the already-decoded
coefficients to the prediction gain to adjust the quantizer resolution
according to the coefficients that are left to decode:
\begin{equation}
Q_{n+1}=\left\lfloor Q_{n}\sqrt{1-\hat{k}^{2}_{n}}\right\rceil \,,\label{eq:quantizer-update}
\end{equation}
where all operations in (\ref{eq:quantizer-update}) are conducted
using bit-exact integer approximations.

\subsection{Residual Coding}

Residual coding is based on Golomb-Rice~\cite{golomb1966run,rice1971adaptive}
codes. The two-sided distribution of the residual is converted to
a one-sided distribution by interleaving the positive and negative
samples using $I\left(x\right)=2\left|x\right|-\mathbb{I}\left(x<0\right)$
as in~\cite{weinberger2000loco}, where $\mathbb{I}\left(\cdot\right)$
is an indicator function.

The Golomb-Rice shift parameter $m$ is coded in the bitstream, as
is the case for FLAC. Unlike FLAC, frames can be broken up into at
most two sub-blocks, with an 8-sample resolution. First, the left
parameter $m_{L}$ is signaled, followed by an interleaved unary code
representing $m_{R}-m_{L}$. If $m_{R}\neq m_{L}$, then the location
of the split is coded. A special value of $m$ is reserved for silence,
allowing for efficient coding of any frame that either starts or ends
with silence. 

\subsection{Pre-emphasis}

\label{subsec:Preemphasis}

The CELT encoder applies the pre-emphasis filter $A\left(z\right)=1-\mu z^{-1}$,
with $\mu=0.85$. Using integer arithmetic, the inverse filter applied
in the decoder can be perfectly reversible, but still has an infinite
impulse response (IIR). Without any additional processing, random
access (seeking) in a file would be \emph{likely} to converge to bit-exactness
within one frame, but due to integer rounding, there is no absolute
guarantee. To overcome the problem, we need to perfectly recover the
last sample of the previous decoded frame (last sample before the
overlap). We can show that any error in the filtering greater than
$\frac{1}{1-\mu}$ will further decay, so the maximum error on the
last sample is $\pm\left\lfloor \frac{1}{1-\mu}\right\rfloor $. Given
that, all we need to do is code the value of the last sample modulo
$K$, with $K=1+2\left\lfloor \frac{1}{1-\mu}\right\rfloor $ guaranteeing
convergence after one frame. Given the CELT pre-emphasis filter, that
corresponds to $K=1+2(6)=13$. In a conventional MDCT-based codec
without pre-emphasis, $K=1$ and no information about the last frame
needs to be transmitted.

\subsection{Inter-Channel Prediction}

For stereo audio, we predict the right channel from the left channel.
We find that prediction in the residual domain results in better compression.
In the encoder, the prediction gain is computed as
\begin{equation}
g=\frac{\sum_{n}e_{L}\left(n\right)e_{R}\left(n\right)}{\sum_{n}e^{2}_{L}\left(n\right)+\epsilon}\,,\label{eq:chan_pred_gain}
\end{equation}
where $e_{L}$ and $e_{R}$ are the prediction residuals for the left
and right channels, respectively, and the small constant $\epsilon$
prevents division by zero. The gain is quantized with a resolution
of $1/16$ and coded with an interleaved Golomb-Rice code centered
around $g=5/16$. In the current version of OLAC, no inter-channel
prediction is applied to the reflection coefficients coding.

\section{Integration and Switching}

\label{sec:Integration-and-Switching}

In both the fixed-bandwidth and ``opportunistic lossless'' cases,
the encoder cannot predetermine if any given frame will ultimately
be encoded losslessly. Instead, it \emph{attempts} lossless coding
of a frame and, if the resulting packet is larger than a threshold
size, falls back to lossy encoding. If the previous frame was encoded
as CELT, then the CELT TDAC and pre-emphasis state must first be copied
to the OLAC state, and vice versa if the previous frame was OLAC.

On the decoder side, the switching is similarly unpredictable. Any
given frame can use a different mode from the previous one. In the
case of a transition from OLAC to CELT, the previous OLAC TDAC/pre-emphasis
state is copied to the CELT state and everything proceeds normally.
In the case of switching from CELT to OLAC, the state is copied in
the other direction and normal decoding proceeds with TDAC and de-emphasis
for the duration of the overlap. Before decoding the overlap period,
the modulo-$K$ value (Sec.~\ref{subsec:Preemphasis}) encoded in
the bitstream is used to force convergence of the de-emphasis filter
state.

Seeking into a fully lossless OLAC file involves a process similar
to mode switching. Because of TDAC and pre-emphasis, attempting to
randomly access samples belonging to frame $i$ requires first decoding
frame $i-1$. Then, when decoding frame $i$, the $\bmod\ K$ value
is used to complete the lossless convergence and all samples from
that point are perfectly lossless.

\section{Evaluation}

\label{sec:Evaluation}

\begin{table*}
\centering{}%
\begin{tabular}{cccccccccc}
\hline 
Genre & \begin{cellvarwidth}[t]
\centering
\textbf{OLAC}

20~ms
\end{cellvarwidth} & \begin{cellvarwidth}[t]
\centering
OLAC 

20~ms w/o

TDAC
\end{cellvarwidth} & \begin{cellvarwidth}[t]
\centering
OLAC

20~ms w/o

TDAC/pre
\end{cellvarwidth} & \begin{cellvarwidth}[t]
\centering
mpeg4als

RM23

-a -o63
\end{cellvarwidth} & FLAC -0 & FLAC -8 & ALAC & \begin{cellvarwidth}[t]
\centering
Monkey's

Audio

-c1000
\end{cellvarwidth} & \begin{cellvarwidth}[t]
\centering
Monkey's

Audio

-c3000
\end{cellvarwidth}\tabularnewline
\hline 
Classical & 42.0 & 42.1 & 43.9 & 42.1 & 45.2 & 42.4 & 45.0 & 42.1 & 41.2\tabularnewline
Jazz & 49.0 & 49.1 & 52.0 & 48.8 & 56.3 & 49.5 & 52.5 & 50.3 & 47.9\tabularnewline
Techno & 58.0 & 58.0 & 60.8 & 57.2 & 66.6 & 58.9 & 61.8 & 60.7 & 57.2\tabularnewline
Rock & 62.5 & 62.4 & 62.7 & 62.1 & 65.8 & 62.5 & 63.5 & 61.5 & 60.4\tabularnewline
Pop & 63.2 & 63.2 & 65.6 & 62.1 & 72.7 & 64.6 & 67.4 & 66.2 & 62.7\tabularnewline
Metal & 66.2 & 66.2 & 68.5 & 64.9 & 73.4 & 67.5 & 70.0 & 67.6 & 65.1\tabularnewline
\hline 
Average & 56.8 & 56.8 & 58.9 & 56.2 & 63.3 & 57.6 & 59.5 & 58.1 & 55.8\tabularnewline
\hline 
\end{tabular}\caption{Compression ratio (lower is better) as a percentage of original size.
OLAC is configured to use 20-ms frames, while all other codecs have
unrestricted frame size. FLAC -0 (-{}-fast) corresponds to the lowest
encoding complexity, whereas FLAC -8 (-{}-best) is the highest complexity.
MPEG-4 ALS is used with the default options, except for the ``-a
-o63'' options that adapt the prediction order up to a maximum of
63 to match OLAC. Monkey's Audio is tested in both ``fast'' (-c1000)
and ``high'' (-c3000) modes. Overall, OLAC is able to achieve 0.8\%
better compression than the best FLAC compression and 2.7\% better
than ALAC, while being 0.6\% worse than MPEG-4 ALS. Compared to Monkey's
Audio, OLAC performs 1.3\% better than ``fast'', but about 1\% worse
than the ``high'' (-c3000) mode. }\label{tab:Compression-ratio}
\end{table*}

\begin{figure}
\centering{}\includegraphics[width=1\columnwidth]{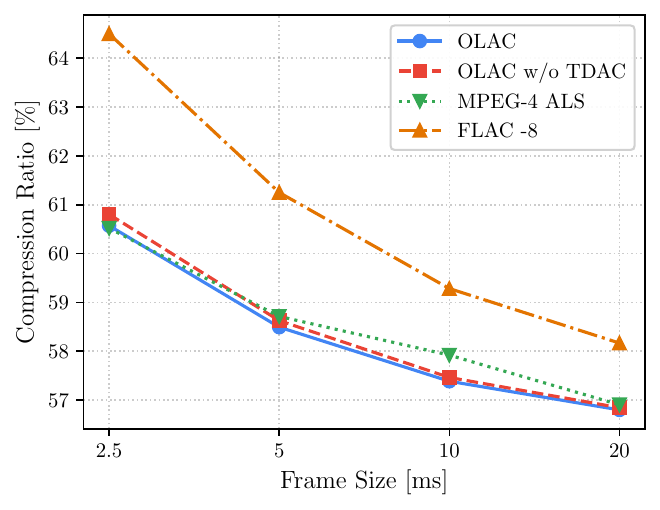}\caption{Effect of the frame size on the average compression ratio (lower is
better) for OLAC, MPEG-4 ALS, and FLAC -8 (-{}-best). For MPEG-4 ALS,
we use options ``-a -o63'' (filters up to order 63) only for 20-ms
frames, since that severely degrades compression on shorter frame
sizes.}\label{fig:Effect-of-frame-size}
\end{figure}

We evaluate OLAC as a standalone codec on a custom corpus of freely
distributable music. The dataset consists of 28 stereo tracks sampled
at 48 kHz in 16-bit PCM format, totaling 105 minutes of audio distributed
across six evaluated genres. We compare OLAC to the popular low-complexity
FLAC and ALAC codecs as well as MPEG-4 ALS and Monkey's Audio. The
results in Table~\ref{tab:Compression-ratio} not only show that
OLAC outperforms FLAC, but also that the use of TDAC has no negative
impact on compression. Moreover, we show that the use of pre-emphasis
improves compression, which we believe is mostly due to improved signal
conditioning resulting in a better filter. The 1\% performance gap
with Monkey's Audio ``high'' setting can easily be explained by
the fact that Monkey's Audio uses a much larger frame size (73,728
samples vs 960 for OLAC) along with more complicated prediction filters
and entropy coding.

Fig.~\ref{fig:Effect-of-frame-size} shows the compression ratio
as a function of the frame size for all frame sizes supported by CELT.
The cost of reducing the frame size from 20~ms down to 2.5~ms is
about 3.6\% for OLAC and MPEG-4 ALS, compared to 6.3\% for FLAC. At
equal frame size, OLAC has on average 0.2\% better compression than
MPEG-4 ALS, confirming that the 0.6\% advantage of MPEG-4 ALS measured
in Table~\ref{tab:Compression-ratio} is only due to the unrestricted
frame size. Moreover, the results confirm that TDAC does not negatively
impact compression (and provides a small benefit), even for very short
frame sizes where the overlap is 100\% (no flat top). 

A C implementation of OLAC is available\footnote{\href{https://github.com/AOMediaCodec/oac/tree/jmvalin/olac_icassp/olac}{https://github.com/AOMediaCodec/oac/tree/jmvalin/olac\_icassp/olac}}
under an open-source license.

\subsection{Complexity}

\begin{table}

\begin{centering}
\begin{tabular}{ccc}
\hline 
Codec & Typical & Worst case\tabularnewline
\hline 
\textbf{OLAC} & 3 & 6\tabularnewline
FLAC & 0.8 & 1.2\tabularnewline
ALAC & 1.5 & 6\tabularnewline
MPEG-4 ALS & 3 & 6\tabularnewline
Monkey's Audio & 12 & 12\tabularnewline
\hline 
\end{tabular}
\par\end{centering}
\caption{Predictor complexity in million MACs per second (MMACS) for each codec
operating at 48~kHz stereo, in the highest compression setting evaluated.
}\label{tab:Predictor-complexity}
\end{table}

The speed at which OLAC runs on a CPU cannot directly be compared
to other codecs since it does not have a highly optimized implementation.
For that reason, we discuss the complexity of the predictor and the
entropy coder, which are the main contributors to the decoder complexity.
Table \ref{tab:Predictor-complexity} shows the predictor complexity
for the different codecs. In this evaluation, OLAC, ALAC, FLAC, and
MPEG-4 ALS all use a similar low-complexity entropy coding model with
Golomb-Rice codes that directly write bits to the bitstream. On the
other hand, Monkey's Audio uses a backward adaptive Laplacian model
with a range coder, resulting in a higher complexity. 

We thus estimate that a suitably optimized OLAC implementation would
be comparable to MPEG-4 ALS in complexity. It would be more complex
than FLAC and ALAC, but significantly less complex than Monkey's Audio
``high'' setting. Considering the use case of a combined lossy/lossless
codec implementation, OLAC must be no more complex than the lossy
codec, which is already achieved by the current unoptimized implementation
alongside CELT.

\section{Conclusion}

In this paper, we have demonstrated that lossless audio coding can
be achieved in a way that is perfectly compatible with MDCT-based
lossy coding, without resorting to a fully integerized MDCT. By combining
a reversible TDAC windowing step via integer lifting with bit-exact
handling of the IIR de-emphasis state, OLAC achieves seamless transitions
without discontinuity or redundant data. Evaluations show that despite
the constraint of matching the CELT overlap structure, OLAC achieves
state-of-the-art compression. It outperforms ALAC and the highest
compression setting of FLAC and remains highly competitive with MPEG-4
ALS and Monkey's Audio, confirming that the use of TDAC and pre-emphasis
introduces no penalty to compression efficiency. Furthermore, the
proposed OLAC codec can be used in real-time applications with short
frame sizes from 2.5~to 20~ms. As a consequence of this work, we
can easily extend most lossy audio codecs to include unified lossy/lossless
compression.

\balance

\bibliographystyle{IEEEtran}
\bibliography{refs}

\end{document}